\documentclass[12pt,preprint]{aastex}
\shorttitle{Spectral regimes of XTE~J1550--564 in its high state}
\shortauthors{Kubota et al.}

\begin{document}

\title{The three spectral regimes found in the stellar black hole
XTE~J1550--564 in its high/soft state}

\author{ Aya {\sc Kubota}}
\affil{Institute of Space and Astronautical
Science, 3-1-1 Yoshinodai, Sagamihara, \\Kanagawa 229-8510, Japan}
\email{aya@astro.isas.ac.jp}
\author{ Kazuo {\sc Makishima}\altaffilmark{1}
\altaffiltext{1}{Also Cosmic Radiation Laboratory,
                 Institute of Physical and Chemical Research }}
\affil{Department of Physics,  University of Tokyo,
7-3-1 Hongo, Bunkyo-ku,\\ Tokyo 113-0033, Japan}

\begin{abstract}
The present paper describes the analysis of
multiple {\it RXTE}/PCA data of the
black hole binary with superluminal jet, XTE~J$1550-564$, acquired
during its 1999--2000 outburst.
The X-ray spectra show features typical of the high/soft spectral state, and
can approximately be described by 
an optically thick disk spectrum plus a power-law tail.
Three distinct spectral regimes, 
named standard regime, anomalous regime,
and apparently standard regime, 
have been found from the entire set of the observed spectra.
When the X-ray luminosity is well below $\sim 6\times10^{38}~{\rm erg~s^{-1}}$
(assuming a distance of 5 kpc),
XTE~J$1550-564$ resides in the {\it standard regime},
where the soft spectral component dominates the power-law component
and the observed disk inner radius is kept constant. 
When the luminosity exceeds the critical luminosity,
the {\it apparently standard regime} is realized, 
where luminosity of the optically thick disk
rises less steeply with the temperature, 
and the spectral shape is moderately 
distorted from that of the standard accretion disk.
In this regime, 
radial temperature gradient of the disk
has been found to be flatter than that of the standard accretion disk.
The results of the {\it apparently standard regime} 
are suggestive of a slim disk 
(e.g., Abramowicz et al. 1988, Watarai et al. 2000) which is a
solution predicted under high mass accretion rate.
In the intermediate {\it anomalous regime},
the spectrum becomes much harder, and
the disk inner radius derived using a simple disk model spectrum 
apparently varies significantly with time.
These properties can be explained as a result of significant 
thermal inverse Comptonization of the disk photons, 
as was found from GRO~J$1655-40$ in its {\it anomalous regime} by
Kubota, Makishima and Ebisawa (2001).

\end{abstract}

\section{Introduction}

In a close binary consisting of a mass accreting stellar-mass
black hole and a mass donating normal star,
the accreting matter 
releases its gravitational
energy as X-ray radiation.
When the mass accretion rate $\dot{M}$ is high, 
such a black hole binary is usually found in a
so-called high/soft state, of which X-ray spectrum is
characterized by a very soft component accompanied by a
power-law tail.
As described, e.g., by Makishima et al. (1986),
the soft spectral component is 
interpreted as thermal emission
from an optically thick accretion disk around the black hole, 
as it can be well reproduced by
a multi-color disk model (MCD model; Mitsuda et al. 1984).
This model approximates a spectrum 
from a standard accretion disk (Shakura \& Sunyaev 1973).
The MCD model has two spectral parameters; 
the maximum disk color temperature,
$T_{\rm in}$, and the apparent disk inner radius, 
$r_{\rm in}$; 
the latter can be related to the true inner radius, $R_{\rm in}$, 
via a correction for spectral hardening (Shimura \& Takahara 1997) and a 
boundary condition (Kubota et al.~1998). 
As the disk luminosity changes significantly,
the value of $R_{\rm in}$ is usually observed to 
remain constant at the innermost Keplerian orbit for the black hole,
$6R_{\rm g}$
where $R_{\rm g}=GM/c^2$ is the gravitational radius
(e.g., Ebisawa et al.\ 1993).

Although this ``standard picture''
is successful for many of the high/soft-state black hole binaries
(e.g., Tanaka \& Lewin 1995; McClintock \& Remillard 2003),
it has been pointed out theoretically 
that the standard disk can be stable only over a limited range of $\dot{M}$.
Actually, a series of new solutions to the accretion flow have been discovered,
including a slim disk solution, which takes advective cooling
into account (Abramowicz et al. 1988; Watarai et al. 2000).
In addition, 
a ``very high state'' has been observationally
found as a derivative state from the soft state
(Miyamoto et al. 1991; van der Klis 1994).
Characterized by an enhanced 
hard component and significant variations in both
$T_{\rm in}$ and $r_{\rm in}$, 
the very high state is regarded as a possible violation of the
simple-minded standard-disk picture. 

A clue to this problem has been recently given by
Kubota, Makishima, \& Ebisawa (2001; hereafter Paper~I)
from an analysis of the multiple {\it RXTE}/PCA data
of the black hole transient with superluminal jet, GRO~J$1655-40$.
While the source behavior was 
described adequately by the standard-disk picture 
over some period (called {\it standard regime}) 
of the entire PCA data span, 
the other period was characterized by the previously suggested 
deviation from such a standard behavior (called {\it anomalous regime}). 
The enhanced hard X-ray spectrum in the {\it anomalous regime}
has been interpreted successfully as a result of 
significant inverse Compton scattering 
of the disk photons 
by some high energy electrons.
The inner radius of the
underlying optically thick disk 
has been found to be kept constant,
when the effect of the Comptonization is taken into account.
This result has been reinforced by Kobayashi et al.~(2003).

In order to reinforce the view obtained from GRO~J$1655-40$, and to deepen our
understanding of the physics of accretion under high values of $\dot{M}$, 
the {\it RXTE} data of the X-ray transient 
XTE~J$1550-564$ is analyzed in this paper.
This transient source 
was discovered on 1998 September 7 by
{\it RXTE}/ASM and {\it CGRO}/BATSE
(Wilson et al. 1998, Smith 1998), and now confirmed as 
a superluminal jet source (Hannikainen et al. 2001).
Figure~\ref{asm} shows a lightcurve of this source,
obtained with the {\it RXTE}/ASM.
As indicated with down-arrows in this figure, 
this outburst was continuously monitored 
by the {\it RXTE} pointing observations through 1999 May 20.
Optical observations have established 
that the system consists of a late type sub giant (G8IV to K4III) 
and a black hole, 
the latter having a dynamical 
mass of $M_{\rm X}=8.4$--11.2 $M_\odot$ (Orosz et al. 2002).
The binary inclination angle and the distance to the source are
estimated to be $i=67^\circ$--$75^\circ$ and
$D=2.8$--7.6 kpc, respectively.
In this paper, $i=70^\circ$ is used as a crude estimate,
and its distance is denoted as $D=D_5 \cdot 5$ kpc.

In \S 2, the observation and data reduction are briefly described.
In \S 3, the PCA spectra are analyzed using
the canonical MCD plus power-law model,
leading to the identification of three characteristic regimes;
{\it standard regime}, {\it anomalous regime}, 
and {\it apparently standard regime}, 
in the increasing order of the luminosity.
It is confirmed in \S 4 that the spectra 
in the {\it anomalous regime}
can be well explained by the strong inverse Compton scattering,
as found in Paper~I.
In \S 5, 
the {\it apparently standard regime} spectra are characterized 
in terms of the radial temperature gradient of the optically-thick disk,
with a conclusion that a slim disk is probably realized.

\section{Observation and data reduction}
\begin{figure}
\plotone{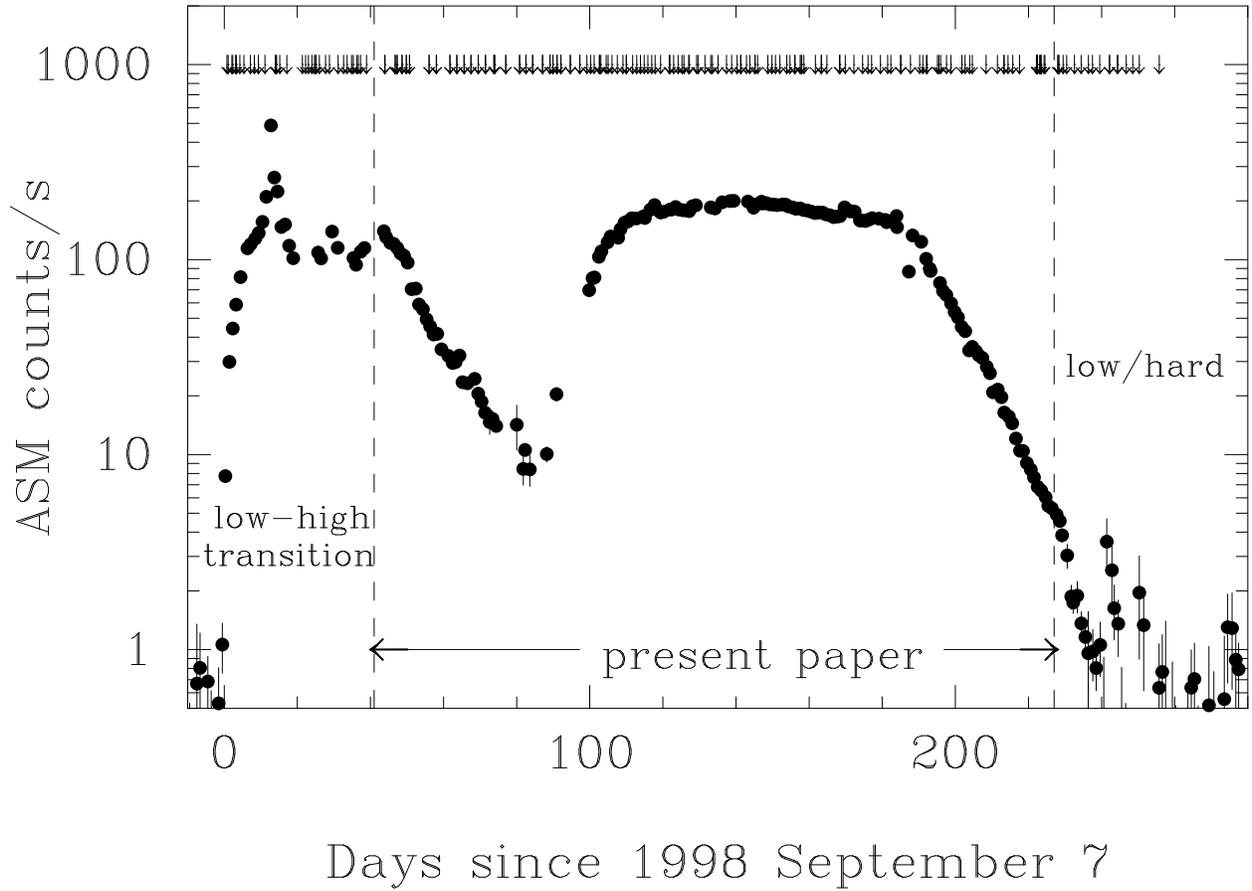}
\caption{The 1.5--12 keV lightcurve of XTE~J$1550-564$, 
obtained with the {\it RXTE}/ASM. 
The pointing observations are indicated with down-arrows.
The horizontal arrow
represents the period studied in the present paper.
\label{asm}}
\end{figure}

As indicated with down-arrows in Fig.~\ref{asm}, the entire outburst was 
covered by 184 pointing observations with {\it RXTE};
the obtained data was analyzed in a standard way
by Sobczak et al.~(2000).
By analyzing the outburst rise phase ($\sim$40 days in Fig~.1) 
covered by 14 pointing observations,
Wilson \& Done (2001) reported that this source 
experienced a significant spectral evolution. 
Figure~1 in their paper shows that 
the power-law photon index describing the PCA 
spectra below $\sim20$ keV switched from 1.4--1.7 to 2--3,
which are typical of the low/hard state (e.g., Tanaka 1997), and 
the high/soft or very high state (e.g., Grove et al. 1998), respectively.
Meanwhile, the 20--100~keV portion of the spectrum maintained a convex shape. 
Therefore, the source is inferred to have made a spectral transition
from the low/hard state into the very high state (Wilson \& Done~2001),
or into the {\it anomalous regime} defined in Paper~I, 
rather than into the canonical high/soft state wherein the
spectrum above 20 keV would not exhibit any cutoff.

These published results indicate that
the spectra acquired during the first $\sim 40$ days 
of the outburst is possibly related to development of
the optically thick accretion disk;
in response to a sudden increase in $\dot{M}$,
an optically thick accretion disk developed inward
and reached the last stable orbit.
Also the data after day 225 (1999 April~20) 
shows characteristics of the low/hard state.
Therefore, the present paper focuses on 
the data taken from 128 PCA pointing observations 
between day 43 (1998 October~20) and 224 (1999 April~19),
which are thought to represent relatively steady states with high $\dot{M}$.
For the systematic analyses of all the data set, 
the HEXTE data is not analyzed here because
of poor statistics in some pointings. 

Following the standard procedure for bright sources,
good PCA data was selected and processed.
The data was excluded 
when the target elevation
angle was less than $10^\circ$ above the Earth's Limb,
or when the actual pointing direction was more than $1^{\prime}\!.2 $
away from the pointed direction.
In particular, the data was discarded 
if it was acquired within 30 minutes
after the spacecraft passage through South Atlantic Anomaly.
The selected data from the individual proportional counter units 
was co-added and used for spectral analyses.
The standard dead time correction procedure was applied to the data.
The PCA background was estimated for each observation,
using the software package {\it pcabackest} (version 2.1e),
supplied by the {\it RXTE} Guest Observer's Facility at NASA/GSFC. 
Sometimes, the use of {\it pcabackest} resulted in an overestimate
of the PCA background up to $\sim10 \%$.
Such a systematic over-estimate was corrected in the same way as in Paper~I.
That is,
the on-source spectra were compared to the predicted model background spectra
in the hardest energy band ($>$80 keV),
where the signal flux is usually negligible.
If necessary, 
the normalization factor of the background spectrum was changed.
The PCA response matrix was made for each observation by utilizing
the software package {\it pcarsp} (version 7.10).
In order to take into account the calibration uncertainties,
1\% systematic errors are added to the data.
Over the 20--35 keV range, the systematic errors are increased to 10\%,
to cope with the response uncertainties
associated with the Xe-K edge at $\sim30$ keV.
Although this could be an overestimate, 
the 20--35~keV data is utilized only in \S4, and 
the results remain essentially unchanged even if it is reduced to 2~\%.

\placefigure{asm}

\section{Standard modeling of the PCA spectra}
\subsection{Characterization of the observed spectra}
The 3--20 keV PCA spectra of XTE~J$1550-564$ were analyzed by  
employing the canonical MCD plus power-law model.
The two constituent continuum components were
subjected to common
photoelectric absorption, with the column fixed at
$N_{\rm H}=1\times10^{22}~{\rm cm}^{-2}$, 
which is reasonable for the source distance of 5~kpc and 
its Galactic position of $(l, b)=(325^\circ\!.9, -1^\circ\!.83)$.
The following results are not affected by changing the value of $N_{\rm H}$ to
$5\times10^{21}~{\rm cm}^{-2}$.
The data requires additionally 
an absorption edge to the power-law component around 7--9 keV
in terms of a smeared edge model (Ebisawa et al. 1994) and a 
narrow Gaussian line around 6.5--6.7 keV for the Fe-K line. 

Figure~\ref{evolution}
shows time histories of the best-fit model parameters,
including the disk bolometric luminosity $L_{\rm disk}$ 
(Mitsuda et al. 1984; Makishima et al. 1986),
the 1--100 keV power-law luminosity $L_{\rm pow}$
calculated assuming an isotropic emission and their sum,
$L_{\rm tot}\equiv L_{\rm disk}+L_{\rm pow}$.
Along the time history, 
the entire observational span can be divided into the following eight periods
(Periods 1--8 denoted in Fig.~\ref{evolution}a)
of relatively distinct properties,
by mainly referring to the 
behavior of $L_{\rm pow}$ and $L_{\rm disk}$.
\begin{itemize}
\setlength{\itemindent}{-10pt}
\item Periods 1 and 7 are characterized by dominance of $L_{\rm pow}$
      and significant reductions in $r_{\rm in}$.
\item In Periods 2 and 8, $L_{\rm tot}$ decreases keeping 
$L_{\rm pow}$ 
rather low, while $r_{\rm in}$ remains rather large and constant.
\item In Period 3, $L_{\rm disk}$ gradually increases,
but $L_{\rm pow}$ remains negligible. 
The power-law component is too weak to constrain its photon index $\Gamma$.
It is therefore fixed at $\Gamma=2.0$.
\item In Periods 4 and 5, $L_{\rm pow}$ is negligible 
(hence $\Gamma$ being still fixed at 2.0), while $L_{\rm tot}$ saturates at 
$\sim 6\times 10^{38}\cdot {D_5}^2~{\rm erg ~s^{-1}}$;
this is about $\sim40\cdot {D_5}^2~\%$ of the Eddington limit, $L_{\rm E}$, 
for a black hole of $10~M_\odot$.

\item Period 6 may be intermediate between Periods 5 and 7.
\end{itemize}
Typical spectra representing these periods are shown in
Fig.~\ref{spec}, and their best-fit parameters are given in Table~1.
The fits are acceptable for the data in Periods 1, 2, and 6--8, 
but sometimes unacceptable in Periods 3--5 (Fig.~\ref{evolution}e). 
This problem is 
considered again in \S5.

\placefigure{evolution}
\placefigure{spec}

\subsection{Observed three spectral regimes}

\begin{figure}	
\begin{center}
\includegraphics[clip=true,width=0.7\textwidth,angle=0]
{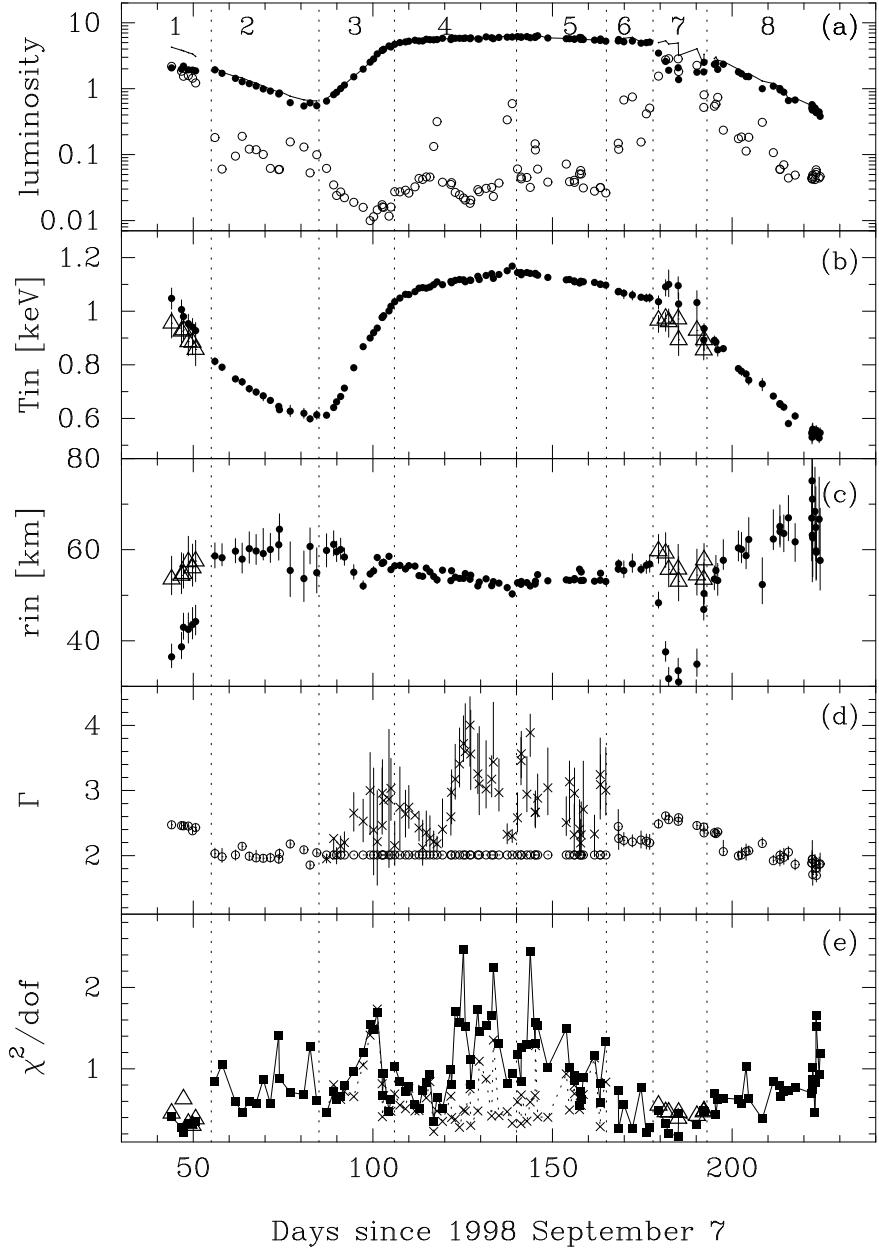}
\caption{
Evolution of the spectral parameters of XTE~J$1550-564$.
(a) Time histories of $L_{\rm disk}$ (filled circles),
$L_{\rm pow}$(open circles),
and $L_{\rm tot}$ (a solid line), all
in the unit of $10^{38}\cdot{D_5}^2~{\rm erg~s^{-1}}$.
As for the disk luminosity, $i=70^\circ$ is assumed.
The eight characteristic periods are indicated at the top.
(b)--(e) Those of $T_{\rm in}$, $r_{\rm in}$, $\Gamma$,
and $\chi^2/{\rm dof}$, respectively.
Open triangles in panel (b), (c), and (e) are
obtained by incorporating the disk-Comptonization.
Crosses in panel (d) and (e) for Period 3--5
show the results from the free-$\Gamma$ fitting.
\label{evolution}}
\end{center}
\end{figure}

\begin{figure}
\begin{center}
\includegraphics[clip=true,width=0.9\textwidth,angle=0]
{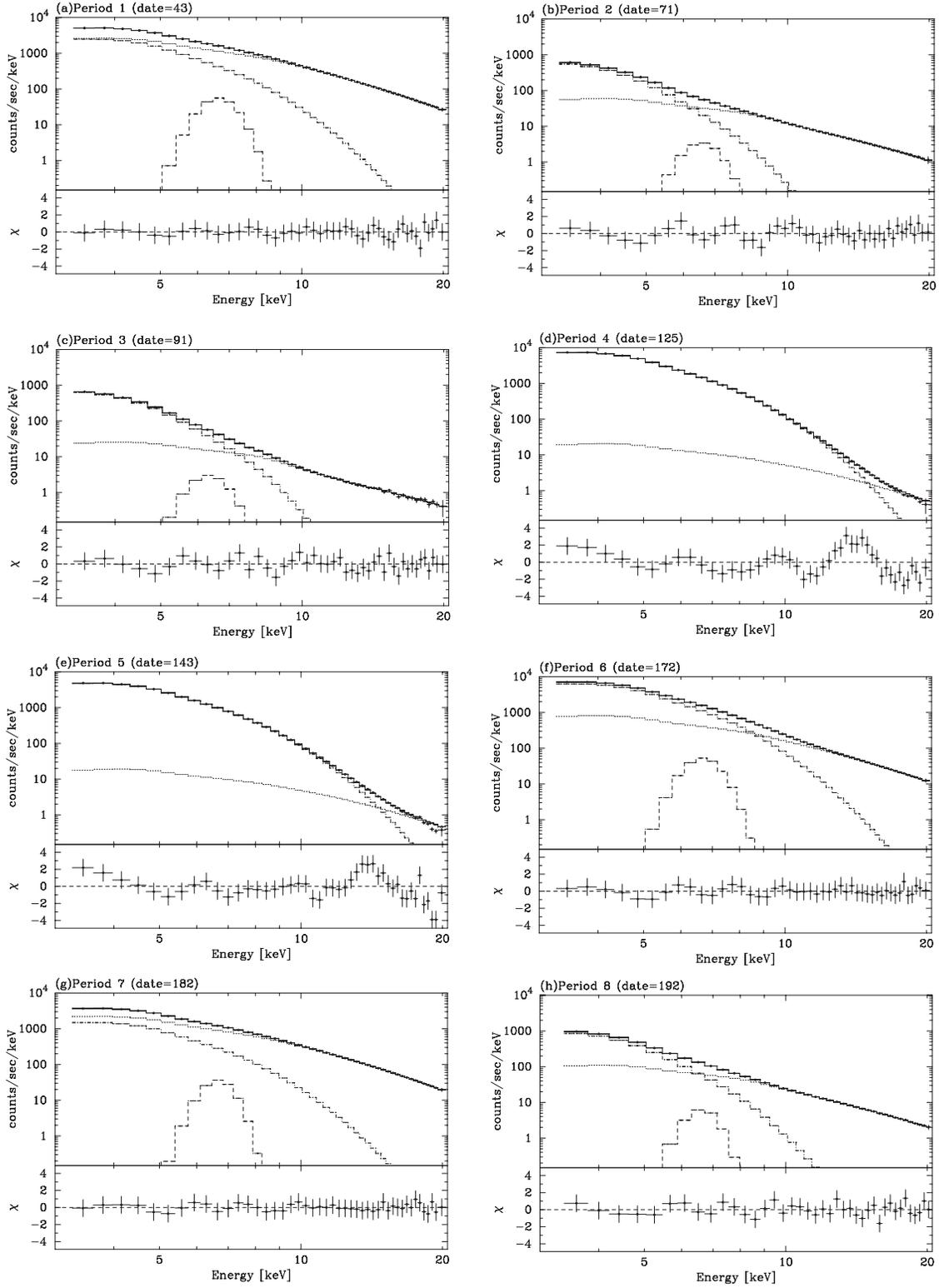}
\caption{Typical PCA spectra of XTE~J$1550-564$ in Periods 1--8, 
compared with the predictions of the
best-fit MCD plus power-law model.
Background has been subtracted, but the instrumental response is not removed.
The lower panel of each spectrum shows fit residuals.  \label{spec}}
\end{center}
\end{figure}

\begin{figure}
\begin{center}
\includegraphics[clip=true,width=0.5\textwidth]
{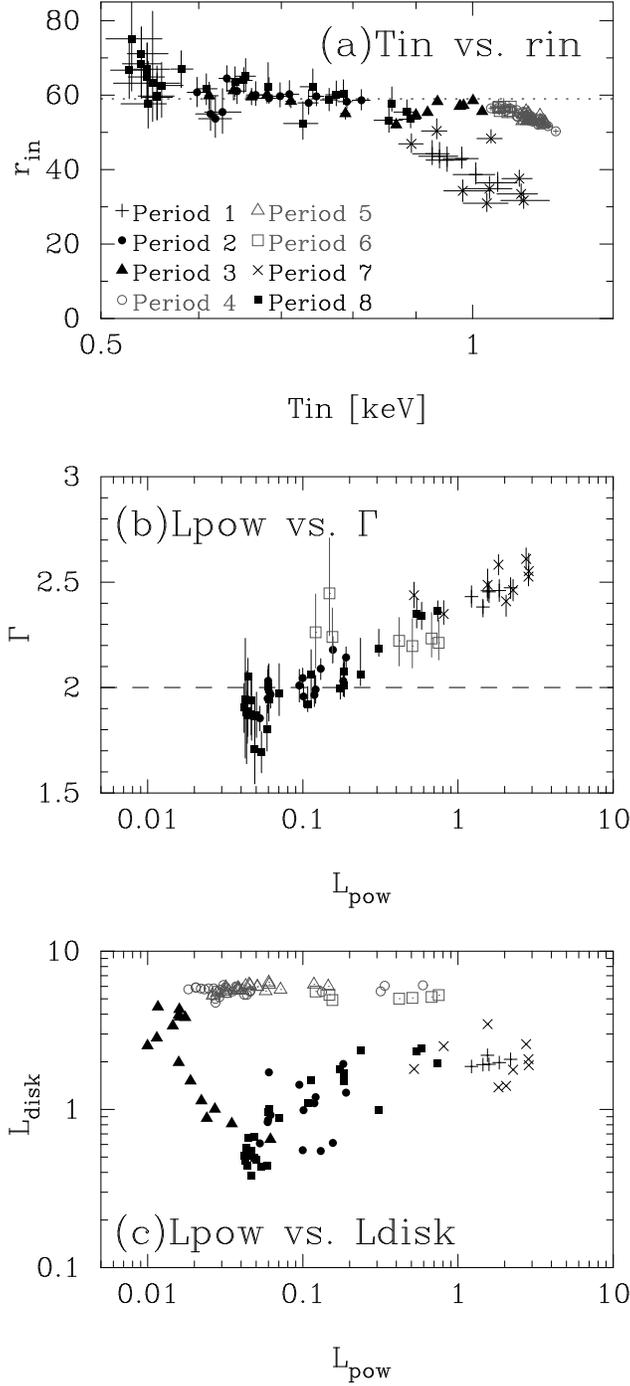}
\caption{Several scatter plots among the spectral parameters.
Panels (a), (b) and (c) shows
$r_{\rm in}$ against $T_{\rm in}$,
$\Gamma$ against $L_{\rm pow}$,
and $L_{\rm disk}$ against $L_{\rm pow}$, respectively.
In panels (a) and (c), Period 1--8 are specified by
eight kinds of symbols presented in panel (a); 
data points of {\it standard}, {\it anomalous} and 
{\it apparently standard regimes} are shown with
filled symbols, crosses, and grey open ones, respectively.
In panel (b), the data for Periods 3--5 are excluded because of the 
weakness of the power-law tail.
\label{scat}}
\end{center}
\end{figure}

To examine each Period for the
validity of the standard picture,
Fig.~\ref{scat} shows several scatter plots 
between the spectral parameters.
The data points of Periods 2, 3, and 8 are thus confirmed to
satisfy the standard picture,
because $r_{\rm in}$ is therein kept constant at $\sim60\cdot D_5$~km
while $T_{\rm in}$ changes significantly, $\sim$ 0.5--1 keV
(Fig.~\ref{evolution}, Fig.~\ref{scat}a).
By taking into account a correction factor for the boundary condition of
$\xi=0.412$ (Kubota et al.~1998) and a color hardening factor of $\kappa=1.7$ 
(Shimura \& Takahara 1997), the true inner radius is estimated as
$R_{\rm in}= r_{\rm in}\cdot \xi \cdot \kappa^2\sim71\cdot D_5$~km. 
This value seems to be slightly smaller than $6R_{\rm g}$ for 
a 10~$M_\odot$ black hole, 90~km, 
though the source distance is not well constrained.

The data points in Period~1 and 7 clearly violate the standard picture.  
Even though the fits are acceptable (Fig.~\ref{evolution}e), 
$r_{\rm in}$ is observed to change
significantly over 30--50$\cdot D_5$~km (Fig.~\ref{evolution}c), and 
$T_{\rm in}$ 
shows strong positive 
deviation from their long-term trend (Fig.~\ref{evolution}b).
In these Periods, the hard component dominates the MCD component (Fig.~\ref{evolution}a), 
and values of $\Gamma$ are largest among these eight Periods 
(Fig.~\ref{evolution}b).
All these properties make Periods 1 and 7 reminiscent of 
the {\it anomalous regime} of GRO~J$1655-40$ (Paper~I).

The classification of Period 4--6 is somewhat ambiguous.
The hard emission is negligible (Fig.~\ref{evolution}a),
and the spectral shape is similar to those in
the {\it standard regime} (Fig.~\ref{spec}).
However, as is clear from Fig.~\ref{evolution}a, the data
suffers from a strong saturation in $L_{\rm disk}$.
In addition, 
a slight increase of $T_{\rm in}$ under the saturation in $L_{\rm disk}$ 
gives rise to a slight decrease in $r_{\rm in}$ (Fig.~\ref{scat}a). 
In this sense, Periods 4--6 are called the {\it apparently standard regime}
in this paper.
Thus, XTE~J$1550-564$ exhibits three characteristic
spectral regimes, the {\it standard} (Periods 2, 3, 8),
{\it apparently standard} (4, 5, 6), and {\it anomalous} (1, 7)
{\it regimes}.

\placefigure{scat}

The $L_{\rm disk}$-$L_{\rm pow}$ diagram, presented in Fig.~\ref{scat}c,
can be used to understand the spectral evolution.
The entire span of this source is found to show
a clockwise loop;
starting from a highest-$L_{\rm pow}$ point (Period 1; {\it anomalous regime}),
the source moves to the left along a low-$L_{\rm disk}$ branch 
(Period 2; {\it standard regime}),
then increases in $L_{\rm disk}$ with $L_{\rm pow}$ kept low (Period 3; {\it standard regime}),
and reaches a ceiling at $L_{\rm disk}$,
$\sim 6\times 10^{38}\cdot {D_5}^2~{\rm erg~s^{-1}}$.
It then moves to the right 
(Periods 4, 5 and 6; {\it apparently standard regime}).
Finally, in Period 7 ({\it anomalous regime}),
it returns to nearly the same position as Period 1,
with Period 8 ({\it standard regime}) being a simple repetition of Period 2.
Thus, the source fortunately exhibited a complete one cycle.

\begin{figure}
\plotone{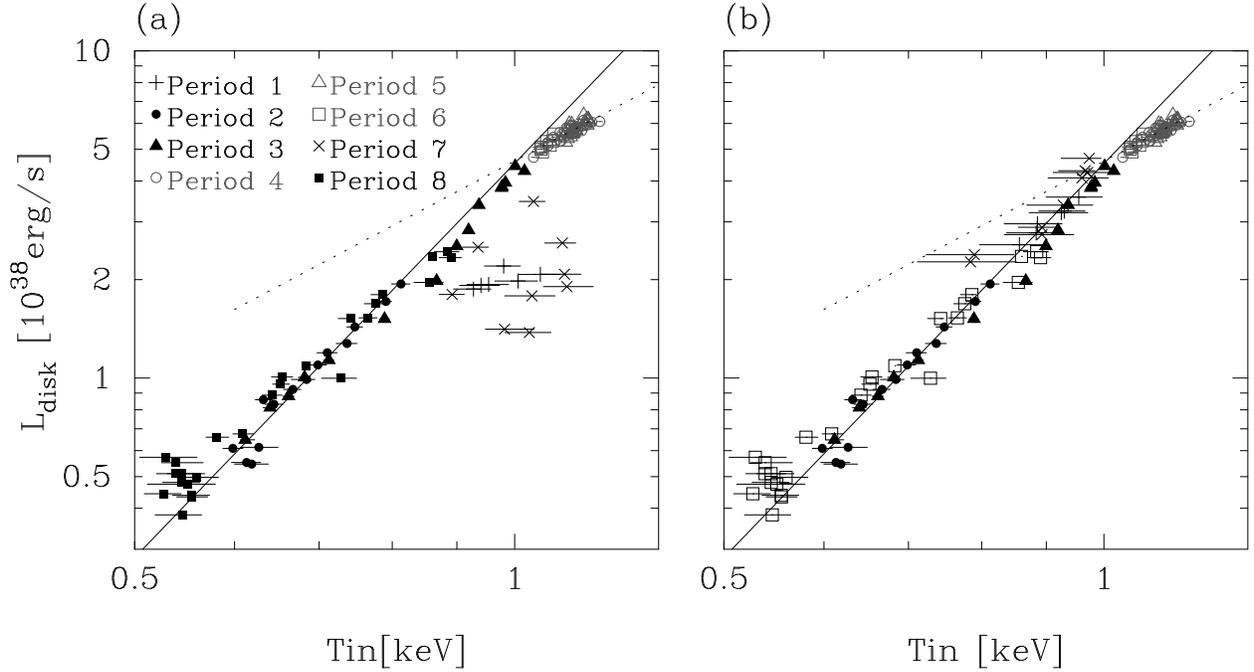}
\caption{(a) The calculated $L_{\rm disk}$ plotted against the observed
$T_{\rm in}$.  Periods 1--8 are specified by the same eight kinds of symbols
as Fig.~\ref{scat}.
The solid and dotted lines
represent the $L_{\rm disk}\propto {T_{\rm in}}^4$ and
$L_{\rm disk}\propto {T_{\rm in}}^2$ relations, respectively.
The source
distance and inclination are assumed to be 5 kpc and $60^\circ$,
respectively.
(b) Same as panel (a), but the data points in Period 1 and 7
({\it anomalous regime}) are
re-calculated considering the Comptonized component as
$L_{\rm disk}+L_{\rm thc}$.
\label{t-l}}
\end{figure}

Figure~\ref{t-l}a shows a $L_{\rm disk}$-$T_{\rm in}$ diagram, 
which is useful to examine the validity of the standard picture
against luminosity.
A simple relation of $L_{\rm disk}\propto {T_{\rm in}}^4$ means 
the constancy of $r_{\rm in}$ (see also Fig~\ref{scat}a), 
and hence the goodness of the standard picture.
In this diagram, the data points in the {\it standard regime}
indeed satisfy this relation, 
while those in the {\it anomalous regime}
deviate significantly. 
The data points in the {\it apparently standard regime},
clustered at the uppermost end of the diagram,
deviate weakly from the standard
$L_{\rm disk}\propto {T_{\rm in}}^4$ relation, exhibiting
a flatter dependence of $L_{\rm disk}$ on
$T_{\rm in}$ as $L_{\rm disk}\propto {T_{\rm in}}^2$.

It is useful to note here that the overall spectral shape of the MCD model 
well describes emission from a standard disk, 
even though it ignores the inner boundary condition of the disk.
As long as the PCA spectra of $T_{\rm in}=0.5$--2~keV are concerned, 
the values of $T_{\rm in}$ and $r_{\rm in}$ 
(with the corrections mentioned before) agree with 4--5~\% with 
those obtained by a more accurate model, e.g., 
a {\sl diskpn} model 
(Gierli$\acute{\rm n}$ski et al. 1999; Gierli$\acute{\rm n}$ski \& Done 2003)
in {\sl xspec}.
Therefore, the characteristics shown in Fig.~\ref{scat}--\ref{t-l}, 
are not due to incompleteness of the MCD model but mean that 
the {\it anomalous regime} and the 
{\it apparently standard regime} 
are intrinsically different from the {\it standard regime}.


\placefigure{t-l}

\section{Reanalyses of the {\it anomalous regime} data\\
-- confirmation of the strong inverse Compton scattering --}

Now that Period 1 and 7 have been inferred to be the {\it anomalous regime},
these data can be reanalyzed employing the concept of the strong
disk Comptonization which successfully explains the
{\it anomalous regime} of GRO~J$1655-40$ (Paper~I).
In the case of GRO~J$1655-40$, 
the hard component in the
{\it anomalous regime} has been found to be different in behavior
from the power-law tail of the {\it standard regime}, 
in several points, including
negative correlation between 
$L_{\rm pow}$ and $L_{\rm disk}$, 
and systematically higher values of $\Gamma$. 
It is hence argued in Paper~I that in the {\it anomalous regime} 
the spectrum is contributed significantly by 
a third spectral component, 
which is harder than the MCD emission
but softer than the power-law tail in the {\it standard regime}.
The strong anti-correlation between $L_{\rm pow}$ and $L_{\rm disk}$
has been taken for evidence that this
third component strongly
and negatively correlates with the MCD component.
It is therefore natural to assume that a fraction of the photons 
from the optically thick accretion disk 
are converted into
the third spectral component, 
most probably through inverse Compton scattering 
by high energy electrons which may 
reside somewhere around the disk.

Following Paper~I, 
the PCA spectra in the {\it anomalous regime} of XTE~J$1550-564$ 
are re-fitted with a three-component model, obtained by adding a Comptonized
component to the original two component model.
In this paper, a thermal Comptonization model
({\sl thcomp}; Zycki, Done, \& Smith 1999 )
is utilized to reproduce the Comptonized component,
instead of the Comptonized blackbody model
({\sl compbb}; Nishimura, Mitsuda, \& Itoh 1986) 
which was used in Paper~I.
The {\sl thcomp} model is based on a solution of the
Kompaneets equation (Lightman \& Zdziarski 1987). 
It well describes thermal cutoff around the electron temperature of the
plasma, $T_{\rm e}$, and
allows the MCD model to be used as a seed photon spectrum.
In contrast, 
the {\sl compbb} model is useable only energies below $kT_{\rm e}$, and it 
assumes a single temperature black body emission 
as a seed photon spectrum.
The {\sl thcomp} model has actually improved the fit goodness compared to 
the {\it compbb} model because of difference in
the seed photon spectra.
The main results of the present paper 
are however independent of such modelings,
and same results were obtained by using the {\it compbb} model.
In this subsection, the
3--50 keV data is used to better constrain the wide-band spectral shape.

For the spectral fitting, many of the model parameters 
were fixed to default values after Paper~I. 
Namely, 
the maximum color temperature of the seed photons 
was tied to $T_{\rm in}$,
to reproduce the situation whereby a part of the
original MCD photons are up-scattered.
Furthermore, $T_{\rm e}$ was
fixed at a representative value of 20~keV
since it could not be constrained and 
is consistent with that obtained during the first 14 observations 
by Wilson \& Done (2001).
Neither a reflection component nor relativistic smearing was added to
the model.
Hence the free parameters of the {\sl thcomp} model are two; 
{\sl thcomp} photon index $\Gamma_{\rm thc}$ which 
expresses the spectral shape below $kT_{\rm e}$, and its normalization.
Moreover, a value of $\Gamma$ of the original power-law component was
fixed at 2.0. 
As a result, the additional number of the free parameters 
for spectral fittings is reduced to only one.

\begin{figure}
\plotone{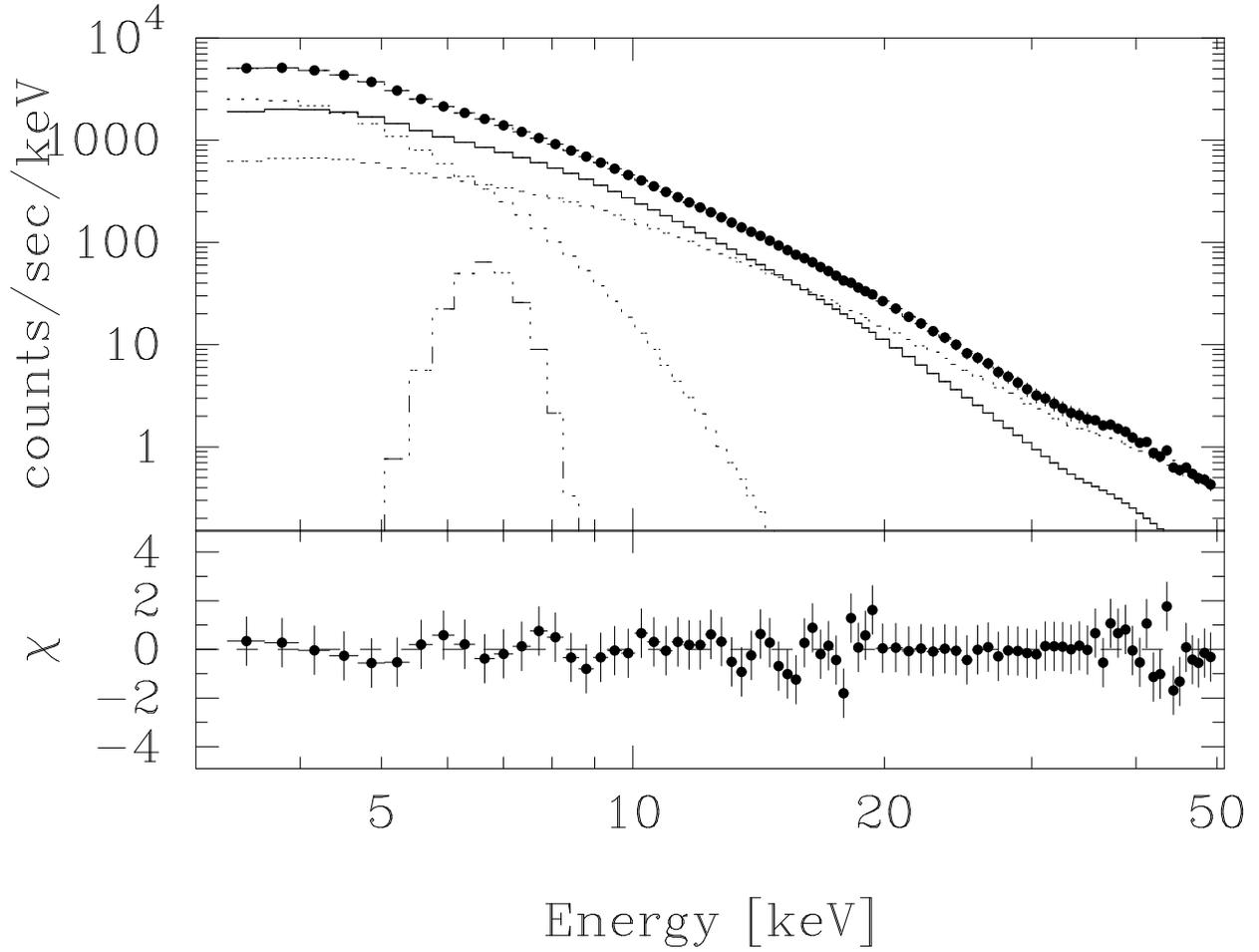}
\caption{The PCA spectrum in the {\it anomalous regime}
(the same as Fig.~\ref{spec}a), fitted with the three-component model 
incorporating the Comptonized component 
(the medium hardness one shown with a solid line).\label{compton}}
\end{figure}

Figure~\ref{compton} shows the same anomalous-regime spectrum
as presented in Fig.~\ref{spec}a, but fitted with
the three-component model over the expanded energy range.
A solid line represents the additional {\sl thcomp} component. 
The fit is acceptable, and the result implies that the dominant 
hard spectral component is 
mostly produced by the strong disk Comptonization.
The re-estimated values of $T_{\rm in}$ for all the data in 
the {\it anomalous regime} are plotted again in 
Fig.~\ref{evolution}b with open triangles.
By considering the disk Comptonization,
the highly deviated data points
in terms of $T_{\rm in}$ have thus settled
back to a smooth long-term trend as was already found in 
GRO~J$1655-40$ (Paper~I).
The luminosity is also re-estimated in Fig.~\ref{t-l}b as
$L_{\rm disk}+L_{\rm thc}$, where $L_{\rm thc}$ is the
estimated 0.01--100 keV {\sl thcomp} luminosity, assuming
an isotropic emission.
Thus, $L_{\rm disk}+L_{\rm thc}$ plotted against the revised $T_{\rm in}$
approximately recovers the standard $L_{\rm disk}\propto {T_{\rm in}}^4$
relation for optically-thick accretion disks.

The value of $r_{\rm in}$ is difficult to be estimated precisely 
under the strong Comptonization, 
because of uncertainties of both geometry and optical depth, $\tau_{\rm es}$,
of the cloud.
In the present paper, it is approximately 
calculated by referring to 
observed photon flux and a re-estimated value of $T_{\rm in}$,
on an assumption that 
few photons are scattered back into the optically thick disk and hence 
the number of the observed 
photons is conserved through inverse Compton scatterings.
Details of this procedure are described in Appendix-A.
The re-estimated values of $r_{\rm in}$, 
plotted in Fig.~\ref{evolution}c with open triangles,
now appear to remain almost stable at $\sim60\cdot D_5$ km.
Consequently, the optically thick accretion disk can also be considered
to remain relatively stable,
even when a significant fraction of the MCD
photons are Comptonized.
Therefore, the picture of disk Comptonization 
is reconfirmed in the {\it anomalous regime},
as suggested in GRO~J$1655-40$ (Paper~I; Kobayashi et al. 2003).

Table~2 summarizes the best-fit parameters associated with 
the exemplified spectra in the {\it anomalous regime}.
The value of $\Gamma_{\rm thc}$ $\sim2.8$
implies a $y$-parameter of
$\sim0.2$, or $\tau_{\rm es}\sim1.8$ 
calculated via a following formula (e.g., Sunyaev \& Titarchuk~1980) 
assuming $T_{\rm e}=20$ keV;
\begin{equation}
\tau_{\rm es}=\sqrt{2.25+\frac{3}{(T_{\rm e}/511{\rm keV})\cdot\{(\Gamma_{\rm thc}+0.5)^2-2.25\}}}-1.5~~~.
\end{equation}
The derived parameters are similar to those of GRO~J$1655-40$. 
The smallness of $y$ and $\tau_{\rm es}$ 
is consistent with the comparatively small fraction of $L_{\rm thc}$
relative to $L_{\rm disk}+L_{\rm thc}$, typically $<0.5$.
It is also consistent with the assumption made in re-evaluating $r_{\rm in}$
in Appendix-A and Fig.~\ref{t-l}b, 
that the mean number of scattering is not too large and 
the fractional energy change is small. 

From these results, 
the anomalous behavior of XTE~J$1550-564$ observed in Periods 1 and 7
can be identified with that of GRO~J$1655-40$ in the {\it anomalous regime}:
the optically-thick standard accretion disk is present,
but the Comptonization converts a significant fraction of its emission
into the hard component.

\section{Reanalyses of the {\it apparently standard regime} data}

\subsection{Properties of the {\it apparently standard regime}}

As shown in \S3, 
the {\it apparently standard regime} 
corresponds to the most luminous phase of the outburst, and 
the data in this regime occupies the upper-right region of the
$L_{\rm disk}$-$T_{\rm in}$ diagram (Fig.~\ref{t-l}b).
In this regime, 
$T_{\rm in}$ gradually changed
keeping $L_{\rm disk}$ almost constant at 
$6\times10^{38}\cdot {D_5}^2~{\rm erg~s^{-1}}$ 
($\sim0.4~L_{\rm E}$).
As a result, the data points deviate from the
standard $L_{\rm disk}\propto {T_{\rm in}}^4$ relation, as 
$L_{\rm disk}\propto {T_{\rm in}}^2$. 
The obtained values of $r_{\rm in}$ are not constant but 
become smaller than those in the 
{\it standard regime}, exhibiting a weak correlation with 
$T_{\rm in}$ as $r_{\rm in}\propto {T_{\rm in}}^{-1}$.
Figure~\ref{scat}a clearly shows this behavior.

As seen in Fig.~\ref{spec}, 
the spectra in the {\it apparently standard regime} 
show a dominant soft component accompanied by a very weak hard tail.
Although these properties are similar to that of the standard soft state,
the {\it apparently standard regime} is something different from the 
{\it standard regime} because of the inconstancy of $r_{\rm in}$ 
(or moderate saturation of $L_{\rm disk}$),
and the absence of the Fe-K line feature that is usually found in the 
{\it standard regime}.
Moreover, the canonical spectral 
model often failed to give acceptable fits to the data 
(Fig.~\ref{evolution}e). 
As are clearly seen in the residuals of Fig.~\ref{spec}d-e,
the discrepancy between the data and the best-fit canonical model appears as 
a low-energy excess and a spectral hump around 13 keV. 
This could be a result of fixing $\Gamma$ of the hard tail component at 2.0; 
accordingly, the fits were repeated by leaving $\Gamma$ free to vary.
Then, the fits became acceptable, but
they required so large values of $\Gamma$ 
(see Fig.~\ref{evolution}d, Table~\ref{tab:mcd}),
that the spectrum below $\sim5$~keV is mostly accounted for by the 
power-law component rather than the MCD component; 
the observed low-energy excess is filled up artificially
by the steep power-law.
Such a fit could be physically inappropriate.

The above results on the {\it apparently standard regime} suggest 
a subtle difference in the accretion disk configuration from the 
standard disk, in such a way that the softest end of the 
observed spectrum is more enhanced than is described by the MCD model.
Such a change, in turn, may arise if, e.g., the radiative efficiency of the 
inner disk region becomes reduced and the radial 
temperature gradient flattens.
In order to quantify this idea, in \S5.2
a $p$-free disk model is constructed 
as a mathematically generalized function of the MCD model. 
The data in the {\it apparently standard regime} is then
examined in \S5.3 by utilizing this model function.

\subsection{Formalism of the $p$-free disk model}
The concept of the standard accretion disk
assumes that the energy released by accretion is
half stored in the Keplerian kinetic energy, and half radiated away as local 
blackbody emission. 
As a result,
the spectrum of the standard disk can be described as a
geometrically-weighted 	
sum of multi-temperature blackbody components
of which the local temperature depends as $T(r)\propto r^{-3/4}$ on the
distance, $r$, from the central black hole.
Therefore, 
any departure of the physical condition assumed by 
the standard disk picture will make the radial temperature gradient 
deviate from the canonical value of $-3/4$. 
Such a deviation will in turn cause
a slight deformation of the
radially-integrated X-ray spectra that are observed.
In order to quantify this idea, 
the MCD model is generalized as seen below,
after initial attempts by 
Mineshige et al. (1994, for GS~$1124-68$) and
Hirano et al. (1995, for Cyg~X-2).

The main assumptions of the model developed here are that
a disk local temperature is described by
$T(r)= T_{\rm in}\cdot (r/{r_{\rm in}})^{-p}$,
and that the disk locally emits a blackbody spectrum.
Here, $p$ is a dimension-less positive parameter
introduced to generalize the MCD formalism, 
with $p=3/4$ implying the MCD model.
The spectrum from this model function can be written as
\begin{equation}
f_p(E) = \frac{2\pi \cos i \cdot r_{\rm in}^2 }{p\cdot d^2}\int^{T_{\rm
in}}_{T_{\rm out}}\left(\frac{T}{T_{\rm
in}}\right)^{-\frac{2}{p}-1}B(E,T)\frac{dT}{T_{\rm in}}~~,
\label{fp}
\end{equation}
with $B(E,T)$ being a blackbody spectrum of temperature $T$.
For convenience, hereafter this mathematical model function is called
$p$-free disk model.
As $p$ decreases, the spectrum becomes softer
than the MCD spectrum of the same $T_{\rm in}$,
because the radiation from outer parts of the accretion disk is emphasized
as seen in equation~(\ref{fp}).

\subsection{Spectral fitting with the $p$-free disk model}

The PCA spectra of Periods 3--5 (one segment of 
{\it standard regime} and two 
of {\it apparently standard regime})
have been re-fitted by replacing the MCD model component with the $p$-free 
disk model.
Although an accurate determination of $p$ is difficult when the
power-law component is strong, the PCA spectra in Period 3--5 are
fortunately free from this obstacle.
The condition of spectral fitting is otherwise the same as in \S 3.
That is, the absorption column $N_{\rm H}$ and
$\Gamma$ of the power-law component are fixed at
$1\times10^{22}~{\rm cm^{-2}}$ and 2.0, respectively.
The Gaussian line was not included.

\begin{figure}
\begin{center}
\includegraphics[clip=true,height=0.7\textheight]
{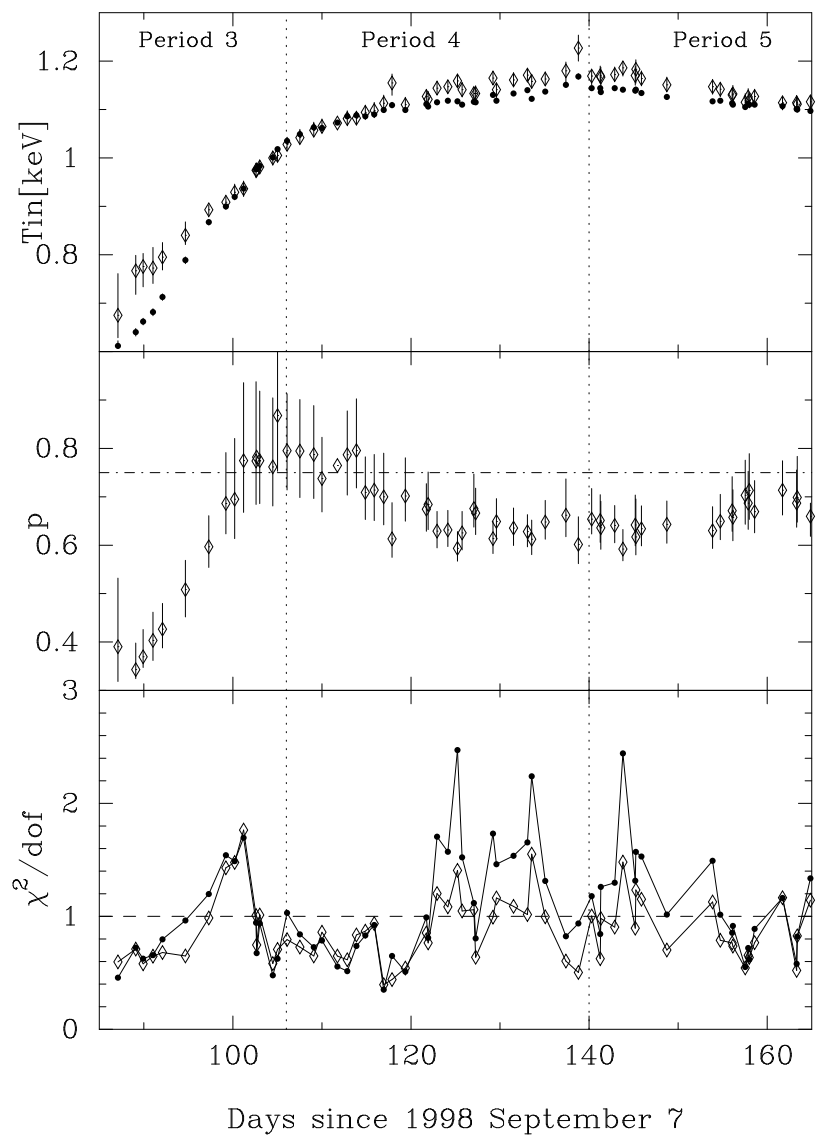}
\caption{Time histories of the best-fit parameters of the $p$-free disk
model (diamond). For comparison, the MCD parameters are also plotted
with filled circles in the top and bottom panels. 
In the middle panel, $p=0.75$ is 
indicated as a dotted line.\label{p-free}}
\end{center}
\end{figure}

\begin{figure}
\plotone{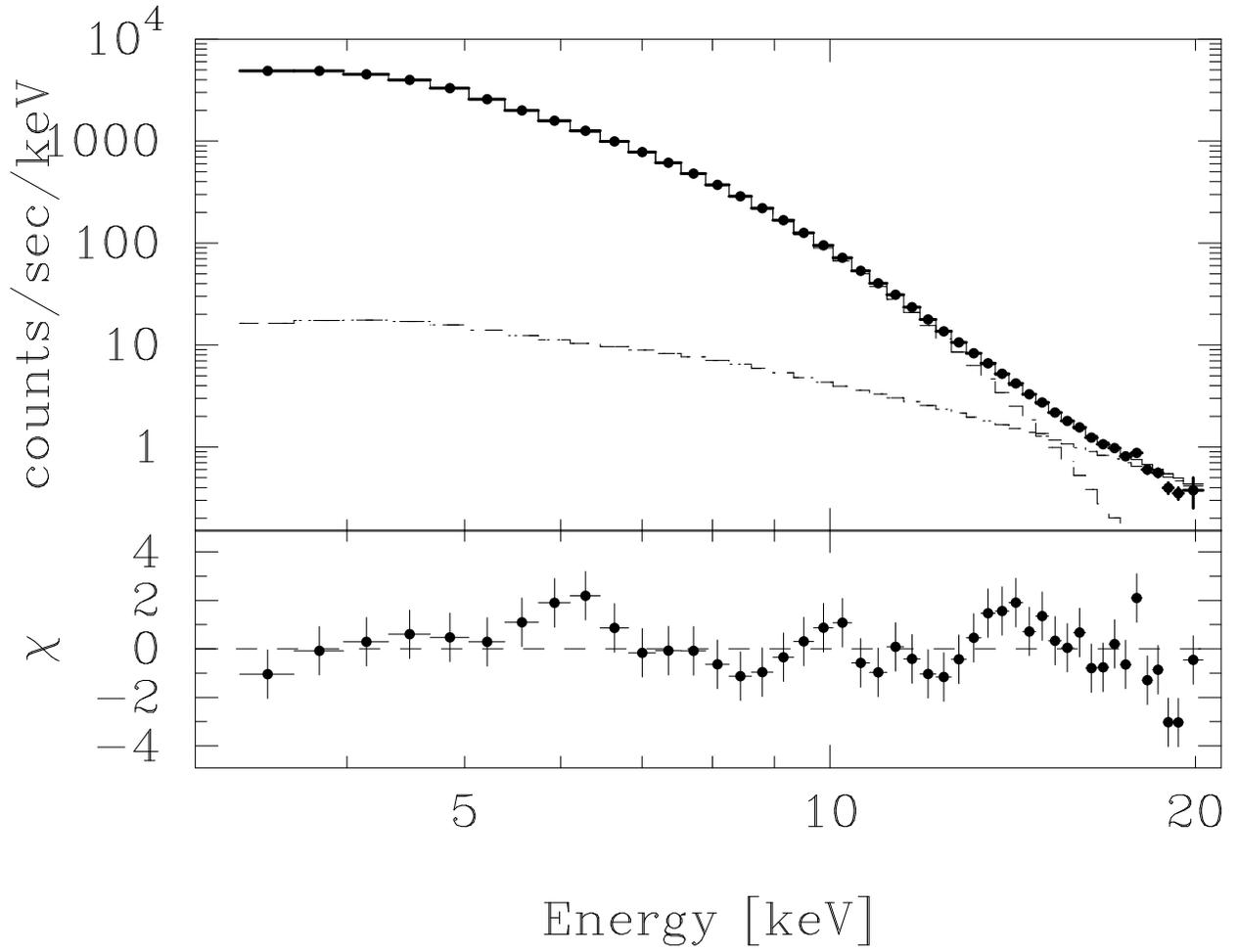}
\caption{The PCA spectrum in the {\it apparently standard regime}
(the same as Fig.~\ref{evolution}e), 
fitted with the $p$-free disk model plus
power-law ($\Gamma=2.0$ fixed).
\label{slim}}
\end{figure}

The time histories of the obtained $p$-free disk model parameters are
given in Fig.~\ref{p-free}. 
In Period 4 and 5 ({\it apparently standard regime}), 
the fit goodness has been significantly improved by allowing $p$ free. 
Figure~\ref{slim} shows 
the same {\it apparently standard regime} spectrum 
originally presented in Fig.~\ref{evolution}e,
fitted this time with the $p$-free disk model, 
and Table~3 shows the examples of the $p$-free model parameters.

\begin{figure}
\begin{center}
\plotone{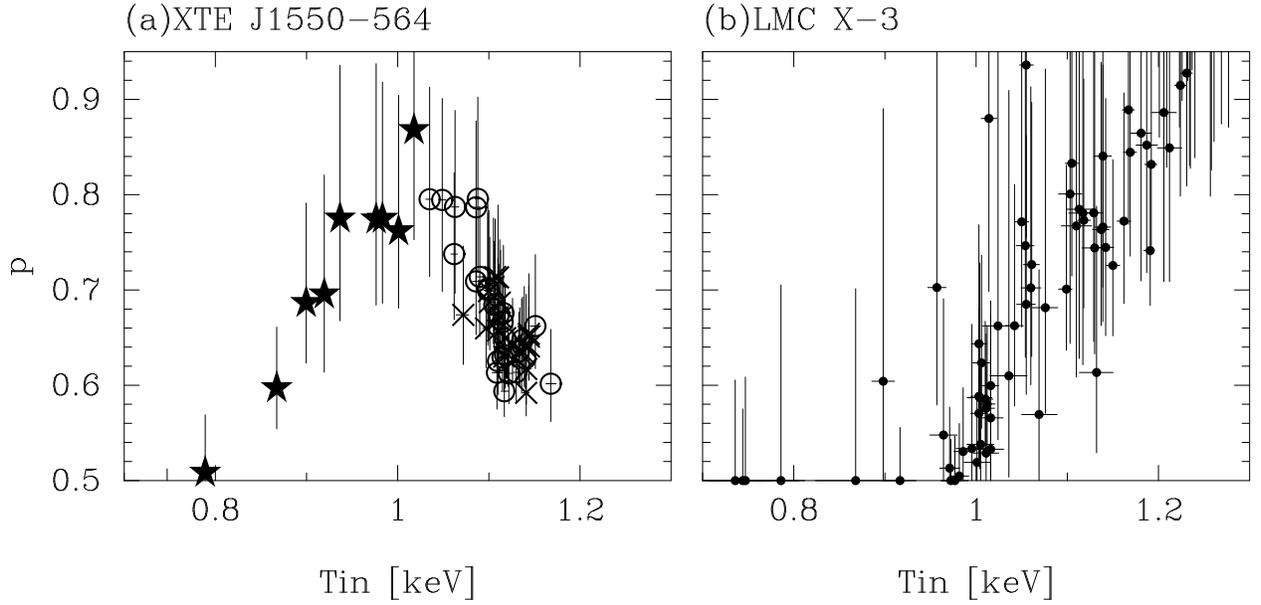}
\caption{The best fit values of $p$, plotted against $T_{\rm in}$ determined
with the MCD model. The PCA 
results for XTE~J$1550-564$ in Period 3--5 (panel a)
are compared with those of LMC~X-3 (panel b). 
Symbols for XTE~J$1550-564$ are the same as in Fig.~\ref{spec}.
In panel (b), data points with large error
bars ($\Delta p>0.5$) are not shown.
As obtained in the MCD fit to the data of LMC~X-3 (Kubota 2001),
the values of $N_{\rm H}$ and $\Gamma$ are
fixed at $7.5\times10^{20}~{\rm cm^{-2}}$ and $\Gamma=2.5$, respectively.
\label{t-p}}
\end{center}
\end{figure}

In Fig.~\ref{t-p}, the best fit values of $p$ are plotted against $T_{\rm in}$;
here, instead of the values of $T_{\rm in}$ obtained
 by the $p$-free disk model, those by the MCD model are
employed, in order to avoid any systematic coupling
between $p$ and $T_{\rm in}$.
The dependence of $p$ on $T_{\rm in}$ thus changes at
$\sim$1 keV. As a function of $T_{\rm in}$, $p$ increases
up to $\sim 1$~keV (Period~3, the {\it standard regime}),
beyond which it decreases abruptly and the correlation turns
negative (Period~4--5, the {\it apparently standard regime}).
To examine these results for various systematic effects, 
the $p$-free disk fits have been extensively repeated by changing the
fitting conditions; 
fixing $N_{\rm H}$ to $5\times10^{21}~{\rm cm^{-2}}$ instead of 
$1\times10^{22}~{\rm cm^{-2}}$, 
fixing $\Gamma$ to 2.2 instead of 2.0, 
including a 
Gaussian line of which the central energy is constrained to 6.2--6.9 keV.
These different conditions slightly affected the absolute values of $p$, 
but did not affect the characteristic $p$ vs. $T_{\rm in}$ behavior of 
Fig.~\ref{t-p}a.

\placefigure{p-free}
\placefigure{slim}
\placefigure{t-p}

Thus, the value of $p$ has been found to deviate from 3/4 as the source 
enters deep into the {\it apparently standard regime}.
However, a still larger excursion of $p$ is observed 
during the {\it standard regime}, where $p=3/4$ should be obtained. 
Therefore, the changes of $p$ in Fig.~\ref{t-p} 
could partially or entirely be due to some artifacts, 
and $p$ could deviate from 3/4,
even if the standard accretion disk is realized.
This could actually happens, 
because the MCD model is a mere approximation of the exact standard-disk
solution; the actual temperature gradient of a standard accretion disk must 
be flatter than $-3/4$ near the innermost disk edge, where
the temperature will approach zero. 
As $T_{\rm in}$ decreases, 
the limited PCA band pass will sample
preferentially the emission from inner disk regions, 
thus making $p$ different from 3/4.


In order to address the above issue, a theoretical approach was first 
attempted. That is, a number of 3--20~keV PCA spectra were simulated, using 
two theoretical model spectra in the {\sl xspec} 
which are known to be more accurate than the
MCD model. One is the {\sl diskpn} model, which 
is based on the Shakura-Sunyaev solution in a pseudo-Newtonian potential;
it properly takes the inner boundary condition into account.
The other is so-called a GRAD model 
(Hanawa 1989; Ebisawa, Mitsuda, \& Hanawa 1991), 
which considers full general relativistic effects for a Schwarzshild metric.
Then, the simulated spectra were fitted with the $p$-free disk model.
As a result, both {\sl diskpn} and GRAD models gave 
$p=0.72$--0.75 as long as $T_{\rm in}$ is in the range 1.2--1.5~keV. 
Furthermore, $p$ was found to change (over 0.5--1.0) for 
lower input disk temperature of $T_{\rm in}=0.8$--1.5~keV. 
However, the $p$-$T_{\rm in}$ relation turned out to be quite
different between the two input models, 
and to depend significantly on the inclination angle when the GRAD model is 
used. In addition to these theoretical complications, neither the 
{\sl diskpn} model
nor the GRAD model 
can be considered accurate yet, since they (as well as the MCD model)
neglect possible spectral deviation from a pure local blackbody. 
Accordingly, this approach has been concluded unrealistic.

In this paper, instead,
the PCA-determined values of $p$ for a standard disk 
has been calibrated empirically, by using actual spectral data of a
prototypical black hole binary, LMC~X-3, in the {\it standard regime}.
This black hole binary perfectly satisfies the standard picture up to 
$\sim L_{\rm E}$ for a black hole of 5--7.2~$M_\odot$
(Paper~I, Wilms et al. 2001), and
its system inclination angle, 65$^\circ$--69$^\circ$, is 
similar to that of XTE~J$1550-564$.
The $p$-free disk model was applied to the PCA spectra of LMC~X-3 
obtained by 128 pointed {\it RXTE} observations from 1996 February to
1999 January (Kubota 2001). 
The best fit values of $p$ from LMC~X-3 are plotted against 
$T_{\rm in}$ in Fig.~\ref{t-p}b.
As expected, a positive 
correlation between $p$ and $T_{\rm in}$ 
artificially appears
even for 
this prototypical ``standard-disk'' object.
Therefore, the behavior of XTE~J$1550-564$ (in Fig.~\ref{t-p}a) 
for $T_{\rm in}\le1$keV (Period~3, the {\it standard regime})
can be understood as an artifact.
In contrast, the behavior of XTE~J$1550-564$ for $T_{\rm in}\ge 1$ keV
is clearly distinct from that of LMC~X-3.
The conclusion is 
that the temperature gradient of XTE~J$1550-564$ in Period~4 and 5 
becomes intrinsically smaller than in the {\it standard regime}, and 
hence the accretion disk in the {\it apparently standard regime} in reality
deviates from the standard picture.

\section{Discussion}
\subsection{Overall picture from the observation}

\begin{figure}
\plotone{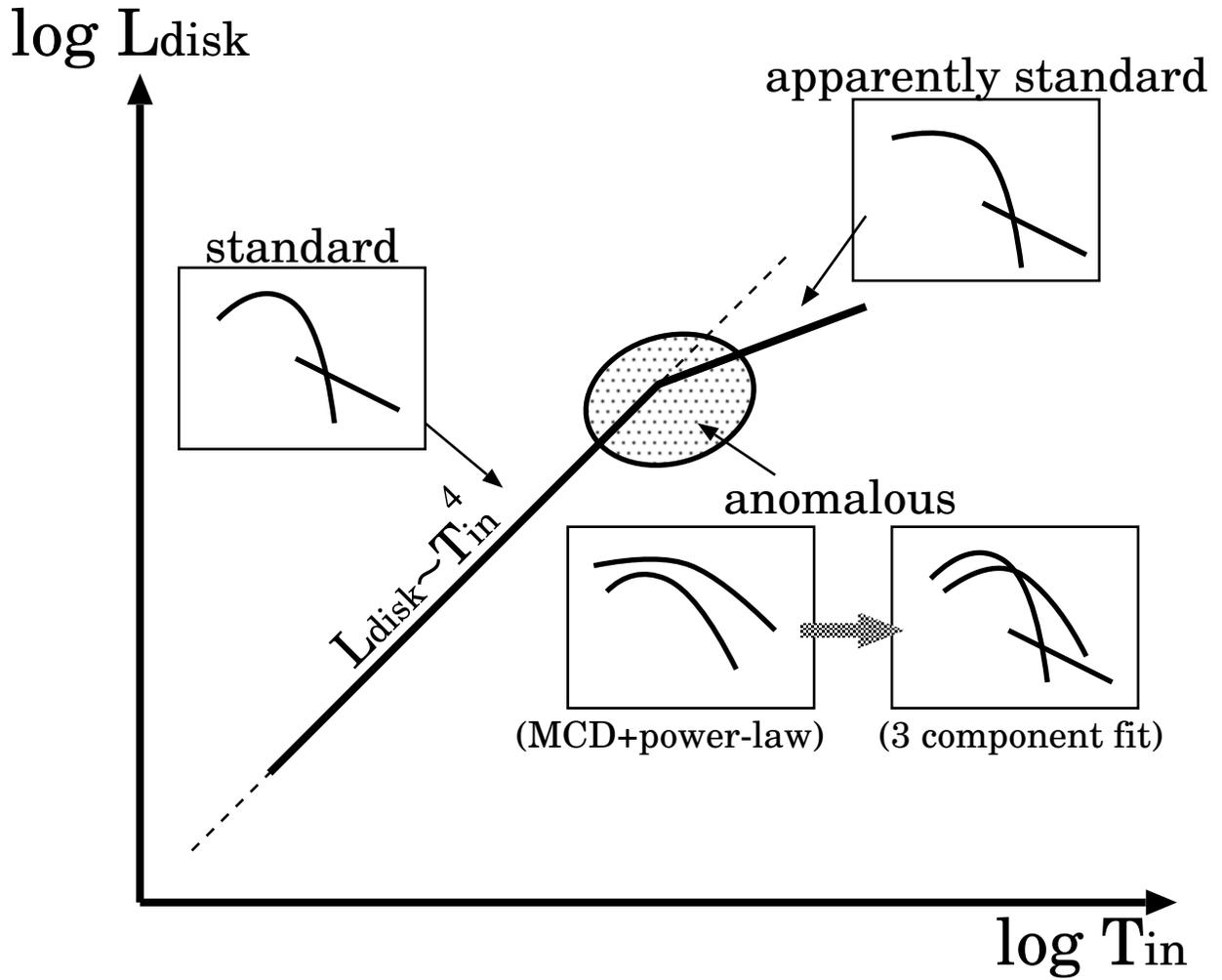}
\caption{
A schematic classification of the three spectral regimes on the 
$L_{\rm disk}$-$T_{\rm in}$ diagram. 
Thick solid and dashed lines show the source behavior obtained under the MCD 
plus power-law fit.
\label{sch}}
\end{figure}

Through the detailed analysis of the PCA data of XTE~J$1550-564$, 
the three spectral regimes have been identified. 
Their relation on the $L_{\rm disk}$-$T_{\rm in}$ plane is illustrated schematically in Fig.~\ref{sch}, while their properties can be summarized as follows.

\begin{enumerate} \setlength{\itemsep}{-1mm}\setlength{\itemindent}{1.8mm}
\item
When $L_{\rm disk}$ is well below a certain critical upper-limit luminosity,
$L_{\rm c}\sim 6\times 10^{38}\cdot {D_5}^2~{\rm erg~s^{-1}}(\sim 0.4\cdot {D_5}^2~L_{\rm E})$, 
the spectral behavior can be explained by the standard-disk picture.
This is the {\it standard regime}.

\item
When $L_{\rm disk}$ hits $L_{\rm c}$, 
it is moderately saturated as 
$L_{\rm disk}\propto {T_{\rm in}}^{2}$, and hence 
$r_{\rm in}$ shows a weak correlation to
$T_{\rm in}$ as 
$r_{\rm in}\propto {T_{\rm in}}^{-1}$.
Although the spectrum, consisting of 
a dominant soft component and a weak hard tail, 
resembles that in the {\it standard regime},
the radial temperature gradient in the disk
(represented by $p$)
becomes flatter than in the standard disk.
This is the {\it apparently standard regime} (\S5).

\item
At an intermediate case ($L_{\rm disk}\sim L_{\rm c}$),
the spectral hard component dominates,
and the apparent inner-disk radius is no longer constant.
This is the {\it anomalous regime} (\S4).
These effects can be explained by 
a sudden increase of the disk Comptonization,
while the underlying disk itself remains in the standard state 
and $r_{\rm in}$ is kept constant.

\item
The source evolves from the {\it standard} to 
{\it apparently standard regimes},
then to the {\it anomalous regime},
and returns again to the {\it standard regime}.

\end{enumerate}
The {\it anomalous regime} 
is naturally identified with that found in GRO~J$1655-40$,
and the scenario of Comptonization suggested in Paper~I successfully 
applies to XTE~J$1550-564$ as well.
It has been confirmed that 
the violent variation in $r_{\rm in}$ in the {\it anomalous regime} 
is apparently caused by strong disk Comptonization, with
the underlying optically thick disk extending down to the last stable 
orbit like in the {\it standard regime}. 
The {\it apparently standard regime}
is possibly the same as Period 1 of GRO~J$1655-40$.
Because intensity variation in Period 1 of the source was small, 
Paper~I did not discuss the source behavior in that period. 
However, 
GRO~J$1655-40$ resides in this period in the upper right corner on the
$L_{\rm disk}$-$T_{\rm in}$ plane (see Paper~I), and its spectra consist of 
a dominant disk component and a very weak hard tail. 
These properties are basically the same as those 
of XTE~J$1550-564$ in the {\it apparently standard regime}.

\placefigure{sch}

\subsection{Comparison with theoretical predictions}

As is well known, theoretical solutions to the steady-state accretion flow
form an $S$-shaped locus
on the plane of $\dot{M}$ vs. the surface density of the disk 
(e.g., Abramowicz et al. 1988, 
Chen \& Taam 1993, Kato, Fukue, \& Mineshige~1998).
The locus involves to thermally stable branches;
the standard Shakura-Sunyaev accretion disk solution, 
and the slim-disk solution, realized when 
$\dot{M}$ is relatively low and very high, respectively.
Evidently, the {\it standard regime} can be identified with the
standard Shakura-Sunyaev solution. 

The slim disk solution takes into account the effect of the advection,
in addition to the viscous heating and the radiative cooling.
This effect becomes important when $\dot{M}$ is high and hence
the luminosity is close to $L_{\rm E}$. 
Under this condition, any increase in $\dot{M}$ would be balanced by an 
increase in the advective transport, accompanied by 
little increase in $L_{\rm disk}$; 
this agrees with the observed mild saturation in $L_{\rm disk}$ 
observed in the {\it apparently standard regime}.
Watarai et al. (2000) simulated many slim disk spectra, fitted them
with the MCD model, and derived an empirical relation between
$r_{\rm in}$ and the apparent $T_{\rm in}$ 
as $r_{\rm in}\propto {T_{\rm in}}^{-1}$ 
(or $L_{\rm disk}\propto {T_{\rm in}}^{2}$).
This is exactly what has been observed in Fig.~\ref{t-l}a.
Furthermore, Watarai et al. (2000) showed that the temperature 
gradient becomes flatter
as the advective cooling becomes important 
because of a progressive suppression of disk emissivity.
In the extreme case, $p$ can reduce to 0.5.
Thus, the overall source behavior in the {\it apparently standard regime}
agrees very well with the prediction by the slim disk model.
Of course, this state assignment is still tentative,
because the slim disk solution still 
neglects important effects such as 
general relativity, magnetic field, and
photon trapping (Ohsuga et al.~2002).
Other solutions would have to be considered as well.

In addition to the two stable branchs, 
a thermally and secularly unstable branch is known to exist between them.
This branch is recognized as a negative slope on the $S$-shape sequence. 
By comparing the obtained $L_{\rm disk}$-$T_{\rm in}$ diagram
to the $S$-shape sequence, the {\it anomalous regime} may have 
some relation to the unstable branch of the sequence.
In other words, the instability of the standard disk may cause the 
{\it anomalous regime}.
Interestingly, the quasi-periodic oscillations (QPOs) are observed 
preferentially in the {\it anomalous regime}, in GRO~J$1655-40$ 
(Remillard et al. 1999) and 
XTE~J$1550-564$ (Remillard et al. 2002). Therefore, the QPO is likely to 
relate to the existence of the Compton cloud.

The observed three distinctive regimes are likely to 
reflect the change of the
accretion disk structure from the standard accretion disk to other solutions, 
as the radiative cooling becomes progressively inefficient.
These results hence provide, at least potentially, 
one of the first observational accounts of the 
long predicted $S$-shaped sequence.




\vspace{1cm}
The authors would like to thank Hajime Inoue, Shin Mineshige and Chris Done,
for their valuable comments. They also thank 
Kazuhiro Nakazawa, Tsunefumi Mizuno 
and Ken Ebisawa for their helpful discussions.
Thanks are also due to Piotr Zycki and Marek Gierli$\acute{\rm n}$ski
for their help with the {\sl thcomp} and the {\sl diskpn} models.
The authors are grateful to Dave Willis for his reading of this paper, and to
the anonymous referee for his/her useful comments.
A.~K. is supported by Japan Society for the Promotion of Science
Postdoctoral Fellowship for Young Scientists.

\appendix
\section{A formula to re-estimate $r_{\rm in}$ under Comptonization}

Under the presence of strong disk Comptonization, 
the apparent disk inner radius $r_{\rm in}$ may be calculated as
\begin{equation}
F^{\rm p}_{\rm disk}+F^{\rm p}_{\rm thc}\cdot 2\cos i= 0.0165 \cdot
\left(\frac{r_{\rm in}^2\cdot \cos i}{ (D/{\rm 10 kpc})^2}\right)\cdot
\left(\frac{T_{\rm in}}{1~{\rm keV}}\right)^3~~~{\rm
photons~s^{-1}~cm^{-2}}~~,
\end{equation}\label{eq:rin}

\noindent
where $F^{\rm p}_{\rm disk}$ and $F^{\rm p}_{\rm thc}$
are 0.01--100 keV photon flux from the direct disk component and the
Comptonized component, respectively.
Here, $T_{\rm in}$ refers to the disk temperature obtained by considering the 
inverse Compton scattering as is done in \S~4.
The first parentheses is just the same as the normalization factor of
the MCD model in the {\sl xspec}.
This formula is based on an assumption that 
there are few photons which are injected again into the
optically thick cool disk 
(i.e., optical depth of the cloud is not very large). 
It can be derived through the following steps.
\begin{enumerate}
\item 
Via flux command of the {\sl xspec}, 
$0.0165 ~{\rm photons~s^{-1}~cm^{-2}}$ in the range of 0.01--100 keV 
is obtained for 
the MCD spectrum of 
which $T_{\rm in}$ and normalization are 1 keV and 1, 
respectively.

\item According to the Stefan-Boltzmann's law, 
photon flux from the original optically thick disk emission is in 
proportion to $r_{\rm in}^2\cdot T_{\rm in}^3$.

\item Isotropic emission is assumed for Comptonized photons, 
and thus $F_{\rm thc}^{\rm p}$ is multiplied by $2\cos i$.

\end{enumerate}

\clearpage
\begin{table}[htbp]
\begin{center}
{\scriptsize
\caption{The best-fit parameters of XTE~J$1550-564$ with the MCD fit}
\label{tab:mcd}
\begin{tabular}{cccccccccr}
\hline \hline
Per.&date $^{\rm a}$&$T_{\rm in}$ [keV]&$r_{\rm in}$[km]$^{\rm b}$&$\Gamma$&$L_{\rm disk}~^{\rm c}$&$L_{\rm pow}~^{\rm c}$ &$E_{\rm c}$ [keV]& eqw [eV]&$\chi^2/\nu$\\
\hline
1 &43&$1.05\pm0.04$&$36^{+3}_{-2}$ &$2.47\pm0.05$&2.07 &2.18&$6.6^{+0.3}_{-0.4}$&$45\pm22$&15.3/37\\
2 &71     &$0.67\pm0.01$&$60\pm3$ &$1.97^{+0.07}_{-0.06}$&0.92&0.062 
&$6.5^{+0.2}_{-0.3}$ &$84\pm28$&21.2/37\\
3 &91    & $0.682\pm0.008$ &$60\pm2$     &(2.0 fixed)&1.01&0.027&$6.3^{+0.3}_{-0.2}$&$80\pm30$&24.9/38\\
 &    & $0.686^{+0.008}_{-0.011}$ &$52^{+2}_{-3}$    & $2.2^{+0.1}_{-0.2}$ &1.02&0.028&$6.2\pm0.3$&$68^{+37}_{-30}$&23.1/37\\
4 &125   &$1.117\pm0.004$&$53.7\pm0.6$  &(2.0 fixed)&5.80&0.022 &---& $<10.3~^{\rm d}$&94.0/38\\
  &    &$1.105^{+0.006}_{-0.004}$  &$54\pm1$&  $3.7^{+0.5}_{-0.3}$&5.73 &0.369&$6.1^{+0.5}_{-0.4}$ &$22\pm17$&18.1/37\\
5 &143 &$1.141\pm0.004$ &$52.1\pm0.6$    &(2.0 fixed)&5.93&0.031&---&$<9.2~^{\rm d}$&92.8/38      \\
 & &$1.133^{+0.007}_{-0.005}$ &$51\pm1$     &$3.9\pm0.3$&5.61&0.834
& $6.1^{+0.5}_{-0.6}$&$18^{+20}_{-15}$    &21.4/37      \\
6 &172     &$1.06\pm0.02$&$57\pm2$&$2.21^{+0.11}_{-0.08}$&5.27    &0.75
&$6.7\pm0.3$ &$46^{+27}_{-11}$  &9.91/37\\
7 &182&$1.10\pm0.05$&$31\pm2$  &$2.55^{+0.08}_{-0.04}$&1.90  &2.85  
&$6.6^{+0.4}_{-0.3}$ &$35\pm19$ &7.5/37\\
8 &192 &$0.78\pm0.02$&$60\pm3$  &$2.01^{+0.11}_{-0.05}$&1.68&0.19
&$6.6\pm0.2$ &$63\pm24$ &17.9/37  \\
\hline
\multicolumn{10}{l}{NOTE.---Errors represent 90\% confidence limits.
The PCA 3--20 keV spectra are used. For all the spectral fitting,}\\
\multicolumn{10}{l}{~~~the {\it smedge} and a Gaussian with fixed  $\sigma=0.1$ keV are included, and $N_{\rm H}$ is fixed at $1\times10^{22}~{\rm cm^{-2}}$.}\\
\multicolumn{10}{l}{$^{\rm a}$~~Days since 1998 September 7.}\\
\multicolumn{10}{l}{$^{\rm b}$~~Apparent inner radii under assumptions of $i=70^\circ$ and $D=5$~kpc.}\\
\multicolumn{10}{l}{$^{\rm c}$~~$L_{\rm disk}$ and $L_{\rm pow}$
represent the bolometric disk luminosity, 1--100 keV
power-law luminosity, }\\
\multicolumn{10}{l}{~~ respectively, 
in the unit of $10^{38}\cdot{D_5}^2~{\rm erg~s^{-1}}$.
As for $L_{\rm disk}$, $i=70^\circ$ is assumed.}\\
\multicolumn{10}{l}{$^{\rm d}$~~ Upper limit 
of Gaussian line with fixed $E_{\rm c}=6.4$ keV.}\\
\end{tabular}
}
\end{center}
\end{table}

\begin{table}[htbp]
\begin{center}
{\scriptsize
\caption{The best-fit parameters with the {\sl thcomp} fit}
\label{tab:thcomp}
\begin{tabular}{ccccccccr}
\hline \hline
Per.&date&$T_{\rm in}$ [keV]&$r_{\rm in}~$[km]$~^{\rm a}$&$\Gamma_{\rm thc}$&$L_{\rm disk}~^{\rm b}$&$L_{\rm pow}~^{\rm b}$ &$L_{\rm thc}~^{\rm b}$&$\chi^2/\nu$\\
\hline
1 &43&$0.96^{+0.04}_{-0.06}$  &$53^{+5}_{-3}$&$2.8^{+0.2}_{-0.1}$
 &2.29 &0.70    &1.27 &33.6/74\\
7 &182&$0.96^{+0.05}_{-0.06}$&$55^{+5}_{-4}$&$2.8\pm0.1$&2.37 &0.69
&1.73 &34.5/74\\
\hline
\multicolumn{9}{l}{NOTE.---The PCA 3--50 keV spectra are used.}\\
\multicolumn{9}{l}{$^{\rm a}$~~Apparent inner radii re-estimated 
by equation (2).}\\
\multicolumn{9}{l}{$^{\rm b}$~~$L_{\rm disk}$, $L_{\rm pow}$, and $L_{\rm thc}$ 
represent the bolometric disk luminosity, 1--100 keV
power-law }\\
\multicolumn{9}{l}{~~luminosity and bolometric (0.01--100 keV) {\sl thcomp}
luminosity, respectively, all in the 
}\\
\multicolumn{9}{l}{~~unit of $10^{38}\cdot{D_5}^2~{\rm erg~s^{-1}}$. Isotropic emission is assumed for $L_{\rm thc}$ and $L_{\rm pow}$, and}\\
\multicolumn{9}{l}{~~ $i=70^\circ$ is assumed for $L_{\rm disk}$.}\\
\end{tabular}
}
\end{center}
\end{table}

\begin{table}[htbp]
\begin{center}
{\scriptsize
\caption{The best-fit parameters with the $p$-free disk fit}
\label{tab:p-free}
\begin{tabular}{ccccccr}
\hline \hline
Per.&date&$T_{\rm in}$ [keV]&$p$&$E_{\rm c}$&eqw [keV]&$\chi^2/\nu$\\
\hline
3 &91    & $0.76^{+0.06}_{-0.02}$ &$0.41^{+0.04}_{-0.05}$ & ---&--- &25.4/39\\
 &    & $0.73\pm0.05$ &$0.5^{+0.2}_{-0.1}$ & $6.3^{+0.3}_{< -0.3}$& 48&22.0/37\\
4  &125    &$1.16^{+0.01}_{-0.02}$  &$0.59^{+0.04}_{-0.02}$ &--- &--- & 54.8/39\\
 & &$1.17^{+0.02}_{-0.01}$  &$0.56\pm0.03$ & $5.9^{+0.2}_{-0.3}$& 45 & 38.9/37\\
5  &143    & $1.18^{+0.02}_{-0.01}$ &$0.60\pm0.03$ &---&---&57.6/39       \\
 & & $1.20\pm0.02$ &$0.57\pm0.03$ &$6.0^{+0.2}_{-0.3}$ &31 &47.7/37       \\
\hline
\multicolumn{7}{l}{NOTE.---The PCA 3--20 keV spectra are used. The power-law photon index is fixed at 2.0.}\\
\end{tabular}
}
\end{center}
\end{table}

\clearpage

\end{document}